\documentclass{osa-article}

\journal{oe}
\articletype{Research Article}
\usepackage{bm}
\usepackage{lineno}
\def\ii{{\mathrm{i}}}
\def\ee{{\mathrm{e}}}

\usepackage{gensymb}
\usepackage{here}
\usepackage{svg}

\begin{document}

\title{Edge-Enhanced Microscopy of Complex Objects using Scalar and Vectorial Vortex Filtering}

\author{Jigme Zangpo,\authormark{*} Tomohiro Kawabe, and Hirokazu Kobayashi.}

\address{Graduate School of Engineering, Electronic and Photonic Engineering, Kochi University of Technology, 185 Miyanokuchi, Tosayamada, Kami City, Kochi 782-8502, Japan}

\email{\authormark{*}jigmezangpo11@gmail.com} %% email address is required



\begin{abstract}
Recently, $4f$ system containing a q-plate has been used to perform edge detection and enhancement of amplitude or phase objects. However, only few studies have concentrated on edge enhancement of complex phase-amplitude objects. Here we experimentally verified the functional difference between scalar and vectorial vortex filtering with the q-plate using an onion cell as a complex object and the vectorial vortex filtering successfully enhanced the edges of phase and amplitude objects in the phase-amplitude object. One problem, however, is indistinguishability of the equally-enhanced edges of the phase and amplitude objects. To address this issue, we propose a method to isolate the edge of the phase object from the edge of the amplitude object using off-axis beam illumination. We theoretically calculated the isolation of the edge of the phase object from the amplitude object, and verified via numerical simulations.
\end{abstract}

%%%%%%%%%%%%%%%%%%%%%%%%%%  body  %%%%%%%%%%%%%%%%%%%%%%%%%%
\section{Introduction}
Traditional bright field microscope (BFM) generates the contrast of an opaque amplitude object (AO) by the absorption of transmitted light in dense areas of the AO\cite{wang2012detecting}. However, BFM is not very useful for transparent phase objects (POs) such as biological cells resulting in low contrast images\cite{khandpur2020compendium}. On the other hand, phase-contrast microscopes (PCMs) provide suitable contrast for PO samples\cite{khandpur2020compendium,burch1942phase}, while this technique cannot highlight the structural details such as edges of samples. PCMs also suffer from phase halos and shade-off in phase-contrast images\cite{frohlich2008phase,lang1982nomarski}. Differential interference-contrast microscopes (DICs) have two significant advantages over PCM and BFM. Firstly, it can detect the edges of the PO, and secondly, samples need not be strained\cite{khandpur2020compendium,frohlich2008phase,lang1982nomarski}. However, DIC detects the edges of samples in only one direction. Hence, the vortex filter employed in $4f$ systems has attracted attention as it is a simple and efficient method to perform isotropic and anisotropic edge-enhanced imaging of biological and medical samples\cite{5,9,11}. This technique has been widely employed in image processing \cite{imageprocessing1,imageprocessing2}, biological imaging \cite{biological1,biological2,biological3}, medical field \cite{medical}, and fingerprint identification \cite{fingerprint1}. 
In general, edge detection is conducted by executing a Hilbert transformation on the object \cite{28,29,30} using an optical vortex filter to yield isotropic or anisotropic edge enhancement\cite{31,32}. 

Recently, studies on edge enhancement of amplitude and phase objects using spiral phase plates have been demonstrated in \cite{spiralPF1,spiralPF2,spiralPF3,spiralPF4,spiralPF5,spiralPF6,spiralPF7,spiralPF8}. Anisotropic edge enhancement has been demonstrated in \cite{biological2,9,11,biological3,spiralPF2,spiralPF5} by changing the topological charge to non-integer values and shifting the center of spiral phase filters. Furthermore, it was revealed that anisotropic edge enhancement can be realized using a superposition of two spiral phase filters\cite{fingerprint1,spiralPF1}.
A spatial light modulator (SLM)\cite{5,11,spiralPF1,spiralPF8} is placed at the Fourier plane to generate a spiral phase filter to enhance the object's edges. However, SLM makes the overall system bulky and limits the resolution \cite{33}. Other vortex filters used in RCP photonics product such as the vortex phase plate (VPP) \cite{RPC1,RPC2} are also used in the Fourier plane for edge enhancement. The vortex filter from SLM and VPP renders the $4f$ system as scalar vortex filtering (SVF) unit. In recent years, vectorial vortex filtering (VVF) has been studied due to its capablility of enhancing phase-amplitude objects (PAOs)\cite{qplate3}. 
VVF can be generated using a space-variant birefringent optical element, q-plate\cite{qplate1,qplate2,qplate3}, spatially variable half-waveplate, and s-wave plate \cite{swaveplate1,swaveplate2} placed at the Fourier plane of a $4f$ system. A q-plate filter is useful as the polarization-sensitive vortex filter to generate both scalar and vectorial vortex filtering\cite{qplate2}. Moreover, q-plates have the advantage of high conversion efficiency of $>97\%$ \cite{34,35,36} compared to other vortex filters. 

In this paper, we experimentally verified the functional difference between scalar and vectorial vortex filtering to enhance the edges of an onion cell as a complex PAO by placing a q-plate in a $4f$ system. 
In an earlier study, the use of q-plates or s-waveplates filter to enhance the edges of simple disk intensity objects\cite{qplate1}, circular aperture \cite{qplate2,swaveplate1}, amoeba \cite{qplate3}, USAF resolution charts as an AO and sprinkled water spots on a glass plate as a transparent PO \cite{swaveplate2} has been studied. Among the studies, Ref.\cite{qplate3} observed the edge enhancement of PAO, but no detailed analysis of PO and AO present in PAO were conducted. We observed PAO using both VVF and SVF, and show that the VVF can highlight the edges of PAO while the SVF results in discontinuity on edge of PO in PAO. Numerical simulation also supports experimental results for VVF and SVF. Despite VVF images offering good edge enhancement, the edges of PO and AO in PAO are indistinguishable because both edges were equally enhanced. Considering this drawback, we calculated and simulated edge isolation of PO from AO using an off-axis beam.

This paper is organized as follows: Section 2 presents the edge enhancement using q-plate mathematically. Both SVF and VVF achieved by q-plate filter are explained. Section 3 presents the experimental results of simple AO, PO, and PAO (biological cell). Next, Section 4 discusses the simulation supporting the theoretical and experimental results of PAO, in addition to discussing how the edge of PO can be isolated from the edge of AO using off-axis beam illumination to the VVF system. 

\section{Theoretical Analysis}
\begin{figure}[ht]
    \centering
    \includegraphics[width=0.80\linewidth]{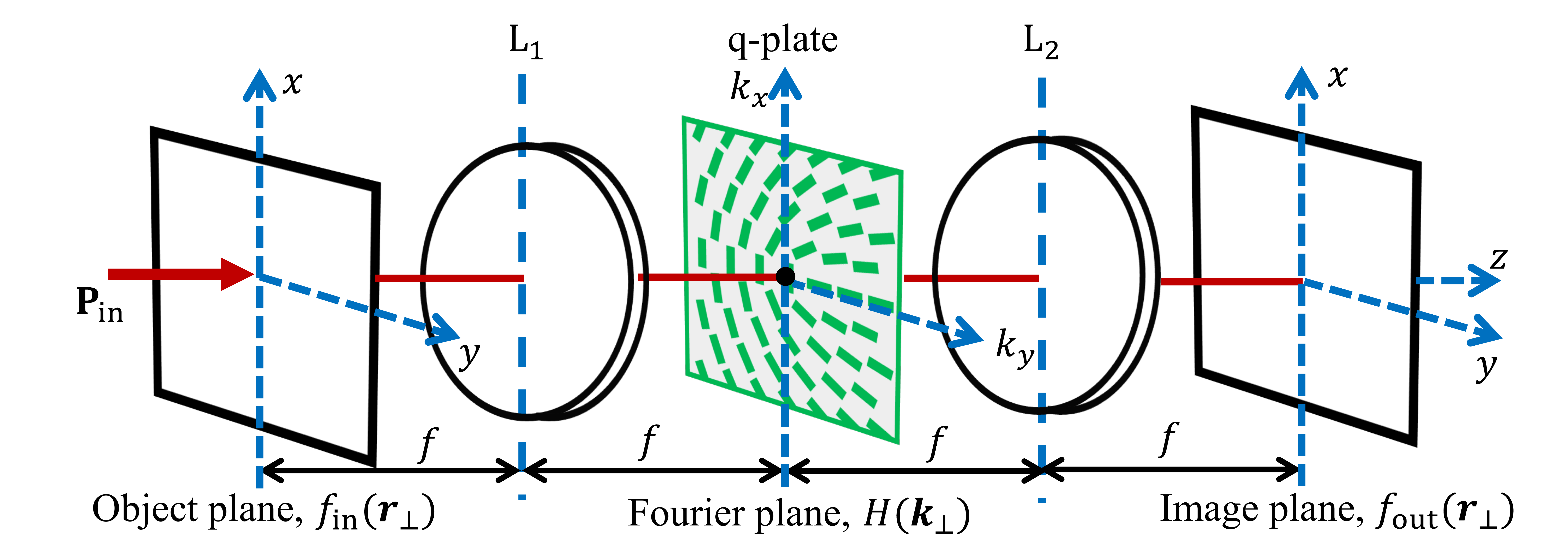}
    \caption{The $4f$ imaging system with on-axis q-plate.}
    \label{fig1}
\end{figure}

The typical BFM technique uses an objective and a tube lens to magnify the sample. To implement all-directional edge enhancement, $4f$ system including a q-plate is placed following the BFM, as shown in Fig. \ref{fig1}. The $4f$ system comprises of three planes: the object, Fourier, and image planes. The object function $f_{\text{in}}(\bm{r})$ is placed on the object plane, the filter function $\bm{H}(\bm{k})$ on the Fourier plane, and the light intensity ${I_{\text{out}}} (\bm{r})$ is observed on the image plane. Here $\bm{r}=(x,y)$ represents the two-dimensional position vector and $\bm{k}=(k_x,k_y)$ is the two-dimensional transverse wavenumber vector. The transmission function of the q-plate\cite{37,38} with retardance $\pi$ radians in polar coordinates is given by the Jones matrix as shown below.

\begin{equation}
    \bm{H}_q(\theta)=\ii\begin{pmatrix}
    \text{cos}(2q\theta)& \text{sin}(2q\theta)\\
    \text{sin}(2q\theta)& -\text{cos}(2q\theta)
    \end{pmatrix},
    \label{Eq1}
\end{equation}
where $m = 2q$ denotes the topological charge, $m$, in a q-plate. The q-plate we use with $q=1/2$ generates light wave with a topological charge $+1$ or $-1$ depending on helicity of circularly-polarized input light wave \cite{35,36} according to Eq. (\ref{Eq1}). The transmission function of the q-plate in Cartesian coordinates derived from Eq. (\ref{Eq1}) in the Fourier domain by setting $q=1/2$ is shown in the following equation.
\begin{equation}
    \bm{H}_{1/2}(\bm{k}) =\frac{\ii}{{k}_{{r}}} 
    \begin{pmatrix}
    k_x& k_y\\
    k_y& -k_x
    \end{pmatrix},
    \label{Eq2}
\end{equation}
where ${k}_{{r}}=\sqrt{\smash{k_x}^2+\smash{k_y}^2}$, and $k_x$ and $k_y$ are the spatial angular frequencies.

The Jones vector of the input polarized light is given by $\bm{P}_\text{in}=\big(\begin{smallmatrix}
  \text{cos}(\alpha)\\\text{sin}(\alpha)e^{\ii\tau}\end{smallmatrix}\big)$, where $\tau$ is the phase difference and $\alpha$ is the azimuth
angle of the state of polarization. The polarized light $\bm{P}_\text{in}$ illuminates the object $f_{\text{in}} (\bm{r})$, and the Fourier transform occurs at $\text{L}_1$. The object spectrum of the Fourier transform is given by $F_\text{in}(\bm{k})\bm{P}_\text{in}$, where $F_\text{in}(\bm{k})$ is the Fourier transform of the input object. When the object spectrum reaches the filter plane, it multiplies with the transmission function of the filter $\bm{H}_{1/2}(\bm{k})$ and gives the modulated spectrum as $F_\text{in}(\bm{k})\bm{H}_{1/2}(\bm{k})\bm{P}_\text{in}$. Then, the modulated spectrum undergoes inverse Fourier transform $F^{-1}$ at $\text{L}_2$ as below:
\begin{equation}
       \bm{f_{\text{out}}} (\bm{r})=F^{-1}\{F_\text{in}(\bm{k})\bm{H}_{1/2}(\bm{k})\bm{P}_\text{in}\}
       =f_{\text{in}} (\bm{r})*\bm{h}(\bm{r}),
       \label{Eq3}
  \end{equation}
where $*$ represents the convolution and $\bm{h}(\bm{r})=F^{-1}\{\bm{H}_{1/2}(\bm{k})\bm{P}_\text{in}\}$, inverse Fourier transform of $H(\bm{k})\bm{P}_\text{in}$, represents the point spread function of the optical system with the q-plate as the Fourier filter.

Equation \ref{Eq3} is simplified further using the Fourier transform property related to differentiation, ${k}_x F_\text{in}({k}_x)\leftrightarrow -\ii\partial_x f_\text{in}(x)$, whereby the image $f_\text{out}(\bm{r})$ is obtained with the convolution term $1/{r}$, where $1/{r}$ is the inverse Fourier transform of $1/{k}_{{r}}$, as follows
\begin{equation}
  \bm{f_{\text{out}}} (\bm{r})=
    \left[(\text{cos}(\alpha)\bm{\nabla} f_\text{in}(\bm{r})+
    \text{sin}(\alpha)e^{\ii\tau}\bm{\nabla_\perp} f_\text{in}(\bm{r})\right] 
    *\frac{1}{{r}},
    \label{Eq4}
\end{equation}
where $\bm{\nabla}=(\frac{\partial}{\partial x},\frac{\partial}{\partial y})$ is the two-dimensional partial differential (gradient) operator on the $(x,y)$-plane and $\bm{\nabla_\perp}=(\frac{\partial}{\partial y},-\frac{\partial}{\partial x})$.
Equation \ref{Eq4} is the general solution for the q-plate filter. 

\subsection{Scalar Vortex Filtering}
When the input light has right or left circular polarization, $\bm{P}_{\text{cir}}=\frac{1}{\sqrt{2}}\big(\begin{smallmatrix}
  1\\ \pm \ii\end{smallmatrix}\big)$, the q-plate works as a single vortex phase element, i.e., SVF. Applying $\bm{P}_\text{cir}$ to Eq. (\ref{Eq4}) and taking absolute square value, the edge-enhanced output intensity, $I_{\text{SVF}} (\bm{r})$, is obtained. By omitting the convolutional term with $1/r$, the edge-enhanced output intensity is approximately calculated as
\begin{equation}
  I_{\text{SVF}} (\bm{r})
    \approx \left|\left(\frac{\partial}{\partial x}\pm\ii\frac{\partial}{\partial y}\right)f_{\text{in}}(\bm{r})
    \right|^2. 
  \label{Eq5}
\end{equation}
By substituting the complex PAO $f_{\text{in}} (\bm{r}) = A(\bm{r})\text{Exp}[\ii B(\bm{r})]$ with the amplitude function $A(\bm{r})$ and phase function $B(\bm{r})$ into Eq. (\ref{Eq5}), the output intensity distribution becomes
\begin{equation}
    I_{\text{SVF}} (\bm{r})= |\bm{\nabla} A(\bm{r})|^2
    +{A(\bm{r})}^2|\bm{\nabla} B(\bm{r})|^2
    \mp 2A(\bm{r})
    \big(\bm{\nabla} A(\bm{r})\cdot \bm{\nabla_\perp} B(\bm{r})\big).
    \label{Eq6}
\end{equation}
In Eq. (\ref{Eq6}), the first and second terms on the right hand side give the edges of the AO and PO, respectively, while the third term on the RHS is a mixture of edges for both amplitude and phase. Depending on the sign of the gradient product $\bm{\nabla} A(\bm{r})\cdot \bm{\nabla_\perp} B(\bm{r})$, the third term will incorrectly enhance or reduce the edges of the AO and PO. 

\subsection{Vector Vortex Filtering}
When the input light has linear polarization, which is superposition of right- and left-circular polarizations, the q-plate works as polarization-dependent vortex filter, i.e., VVF. By substituting $\bm{P}_{\text{lin}} =\big(\begin{smallmatrix} 1\\0\end{smallmatrix}\big)$ into Eq. (\ref{Eq4}), taking absolute square value, and omitting the convolution term, the edge-enhanced output intensity, $I_{\text{VVF}} (\bm{r})$, is obtained as follows.
\begin{equation}
  I_{\text{VVF}} (\bm{r})
    \approx |\bm{\nabla} f_{\text{in}} (\bm{r})|^2.  
  \label{Eq7}
\end{equation}
By substituting the same complex PAO, $f_{\text{in}} (\bm{r})$, used in the SVF into Eq. (\ref{Eq7}), the following output intensity is obtained.  
\begin{equation}
  I_{\text{VVF}} (\bm{r})=|\bm{\nabla} A(\bm{r})|^2+|A(\bm{r})\bm{\nabla} B(\bm{r})|^2.  
  \label{Eq8}
\end{equation}
 As Eq. (\ref{Eq8}) shows, both the amplitude and phase edges can be enhanced and are separated from each other's edges. Unlike the case of SVF, there is no other term contributing to the AO and PO edges. 

\section{Experimental setup and Results}
\begin{figure}[ht]
    \centering
    \includegraphics[width=0.90\linewidth]{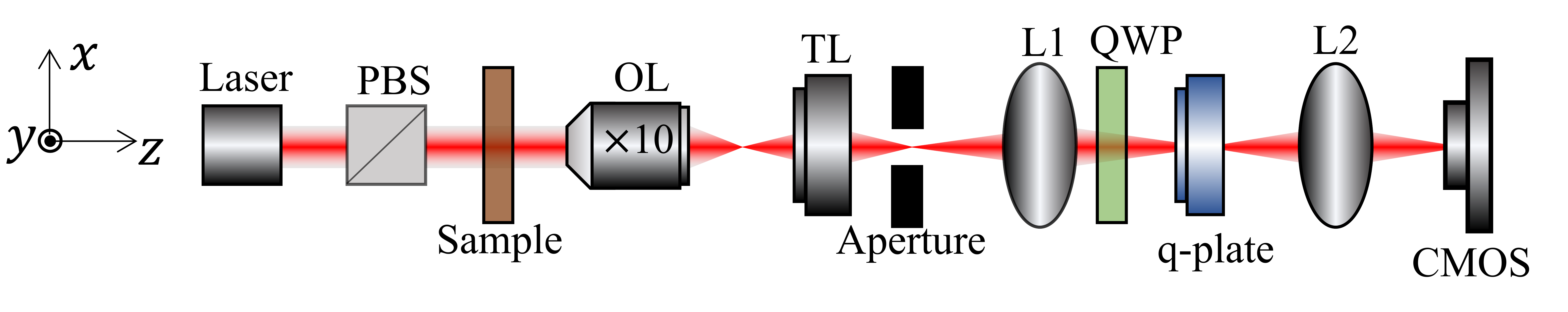}
    \caption{The sketch of the experimental setup. PBS is Polarizing Beam Splitter; OL is objective lens; TL is tube lens; L1 and L2 are Fourier lenses; QWP is quarter wave plate; CMOS is the Complementary Metal-Oxide-Semiconductor camera. $f_{\text{OB},\text{TL},1,2}=18,180,300,250 \text{mm}$, $f_{\text{OB}}$ and $f_{\text{TL}}$ are focal lengths of OL and TL. $f_\text{1}$ and $f_\text{2}$ are the focal lengths of L1 and L2.}
    \label{exp}
\end{figure}

Figure \ref{exp} shows an experimental setup of a $4f$ imaging system with a q-plate to enhance the edge of an object. The setup can serve for both VVF and SVF. 

The first $4f$ system is typical BFM, comprising an objective lens (focal length, $f_{\text{OB}}=18$mm) with a magnification factor of 10 and a tube lens ($f_{\text{TL}}=180$mm). The second $4f$ system includes a q-plate (retardance $\pi$, and topological charge $1$) located at the Fourier plane, between the midpoint of rear focal length ($f_1=300$mm) of Lens $1$ and front focal length ($f_2=250$mm) of Lens $2$. We used commercially available q-plates identified as zero-order vortex half-wave retarders (Thorlabs, WPV10L-633). A quarter waveplate (QWP) known as zero-order quarter-wave plate (Thorlabs, WPQ10M-633) was placed before the q-plate. At the image plane, CMOS camera (Thorlabs, DCC1645C) was placed.  The CMOS camera  has a imaging area $4.6 \text{mm}\times3.7 \text{mm}$, array format $1280\text{H}\times1024\text{V}$, and pixel size of 3.6 \textmu m$\times$3.6 \textmu m. 

The illumination beam is provided by a laser diode (Thorlabs, HLS635) with a wavelength $635$ nm. A polarizing beam splitter (PBS) is used to produce the horizontal linearly polarized light to illuminate the sample. The laser light through the object passes through the first $4f$ system. Then, the beam carrying the magnified object incident on the second $4f$ system. When a magnified object is transmitted via the second $4f$ system, it is Fourier transformed at Lens $1$ and its spectrum is multiplied with the q-plate function at the Fourier plane. Then, the modulated object spectrum undergoes inverse Fourier transform at Lens $2$ and the image is recorded by a CMOS camera. The image obtained is the reconstructed image of the target object after spatial filtering, where only the edges are highlighted.
When the QWP in the $4f$ system is placed at 0\degree, the input light becomes linear polarization and system becomes VVF. On the other hand, in the case of the QWP with 45\degree, the input polarization becomes circular one and system become SVF.
We placed a circular aperture at the object plane of the second $4f$ imaging system to increase the resolution of POs.

\subsection{Experimental Results of simple AO and PO}
\begin{figure}[ht]
    \centering
    \includegraphics[width=0.40\linewidth]{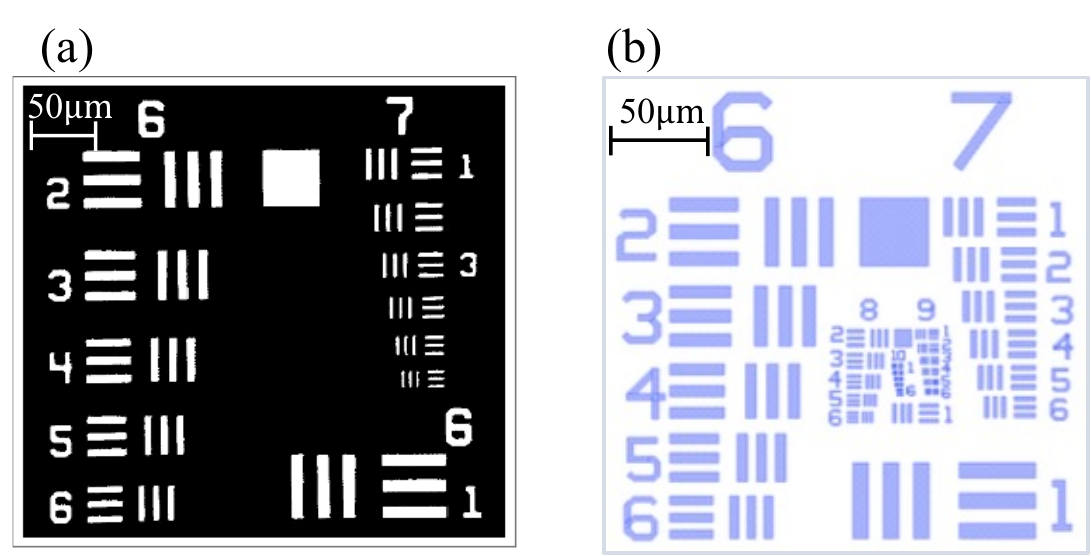}
    \caption{Simple Target object, (a) R1DS1N-Negative 1951 USAF Test Targets, amplitude object and (b) Quantitative Phase Target, phase object.}
    \label{obj}
\end{figure}

To demonstrate our theory, we used commercially available AOs from Thorlabs, Inc. called R1DS1N-Negative 1951 USAF Test Targets and a PO from Benchmark Technologies identified as Quantitative Phase Target (refractive index of 1.52). We used distinct feature heights of the PO of 100, 150, 200, 250, 300, and 350 nm with the phase difference ranging from 0.56 to 1.99 radian at the wavelength of 635 nm. The AO and PO are shown in Figs. \ref{obj} (a) and (b), respectively. 

\begin{figure}[ht]
    \centering
    \includegraphics[width=0.90\linewidth]{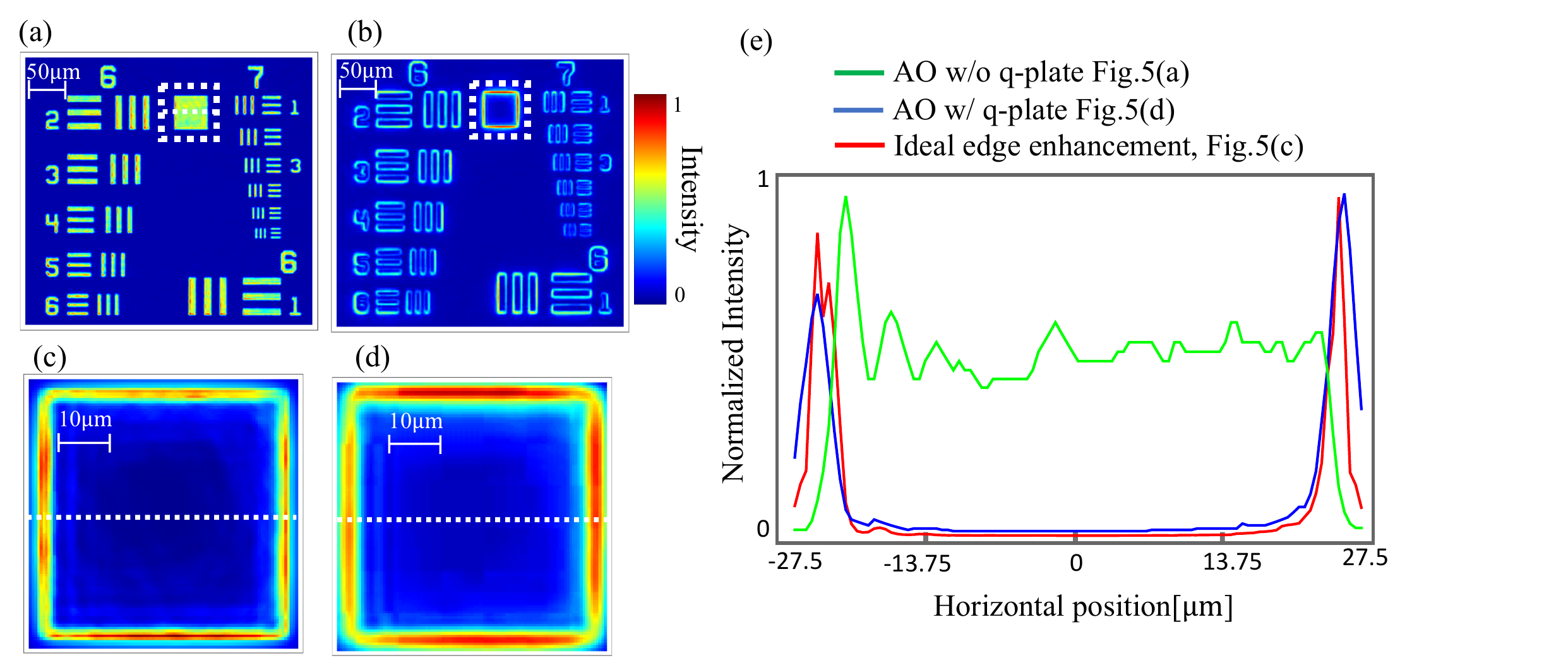}
    \caption{(a) and (b) Edge enhancement without and with q-plate, respectively, (c) ideal edge enhancement, (d) extracted portion of square from Fig. \ref{1exp} (b), and (e) horizontal cross-sectional intensity distribution along white dashed line of Figs. \ref{1exp} (a) (green curve),  (d) (blue curve), and (c) (red curve).}
    \label{1exp}
\end{figure}

First, the AO was used to conduct the experiment. For an easy comparison, the q-plate was removed first to obtain the object image, as shown in Fig. \ref{1exp} (a). Then, the q-plate was inserted at the Fourier plane of the $4f$ imaging system in the experimental setup to obtain the object image, as shown in Fig. \ref{1exp} (b). 

For a quantitative comparison of the quality of the filtered image, the square portion in Fig. \ref{1exp} (a) was extracted and numerically simulated with the q-plate for ideal edge enhancement, which is shown in Fig. \ref{1exp} (c). Similarly, the wrapped square portion in Fig. \ref{1exp} (b) was extracted and shown in Fig. \ref{1exp} (d). The correlation coefficient between Figs. \ref{1exp} (c) and (d) is 0.919, which indicates that edge enhancement using q-plate is almost ideal. The horizontal cross-sectional intensity distribution plotted in \ref{1exp} (e) has green, red, and blue curves that correspond to the pixel-value of the white dashed line of Figs. \ref{1exp} (a) (wrapped square portion), (c) and (d), respectively. The green curve in Fig. \ref{1exp} (e) shows that the edge of the object is not detected when no q-plate is used, compared to the blue curve when a q-plate is used. According to Fig. \ref{1exp} (e), it also suggests edges of experimental results with q-plate, blue curve, and ideal edge, red curve are strongly correlated.  However, the edge is slightly wider in the experimental result with the q-plate than the ideal edge, and the intensity inside the square is less in the ideal edge than that in the experimental result. This is due to the presence of the convolution term ($1/r$, refer Eq. (\ref{Eq4})) in the experimental result, which is not present in ideal edge enhancement. Some of the distinctive attributes for good edge enhancement are sharper edge intensity and intensity approximately zero inside the object.

\begin{figure}[ht]
    \centering
    \includegraphics[width=0.90\linewidth]{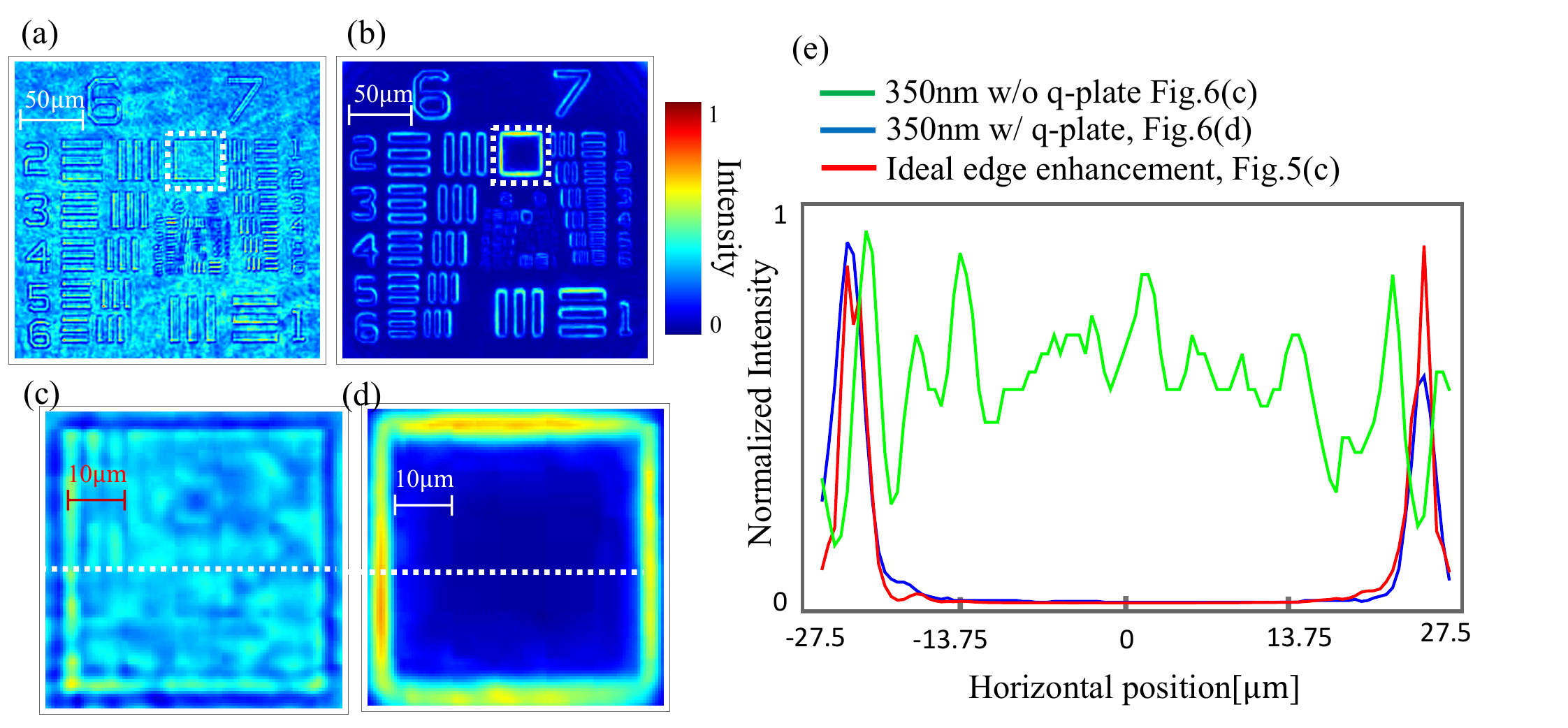}
    \caption{(a) and (b) Edge enhancement of PO 350-nm without and with a q-plate, (c) extracted portion of square from Fig. \ref{2exp} (a), (d) extracted portion of square from Fig. \ref{2exp} (b) and (e)
    horizontal cross-sectional intensity distribution along white dashed line of Fig. \ref{2exp} (c) (green curve), Fig. \ref{2exp} (d) (blue curve), and Fig. \ref{1exp} (c) (red curve).}
    \label{2exp}
\end{figure}

Secondly, the PO was used as a target. The experimental results shown in Figs. \ref{2exp} (a), and (b) are the edge enhancement of PO (height $350$ $\text{nm}$) without and with the q-plate, respectively.

\begin{figure}[ht]
    \centering
    \includegraphics[width=0.60\linewidth]{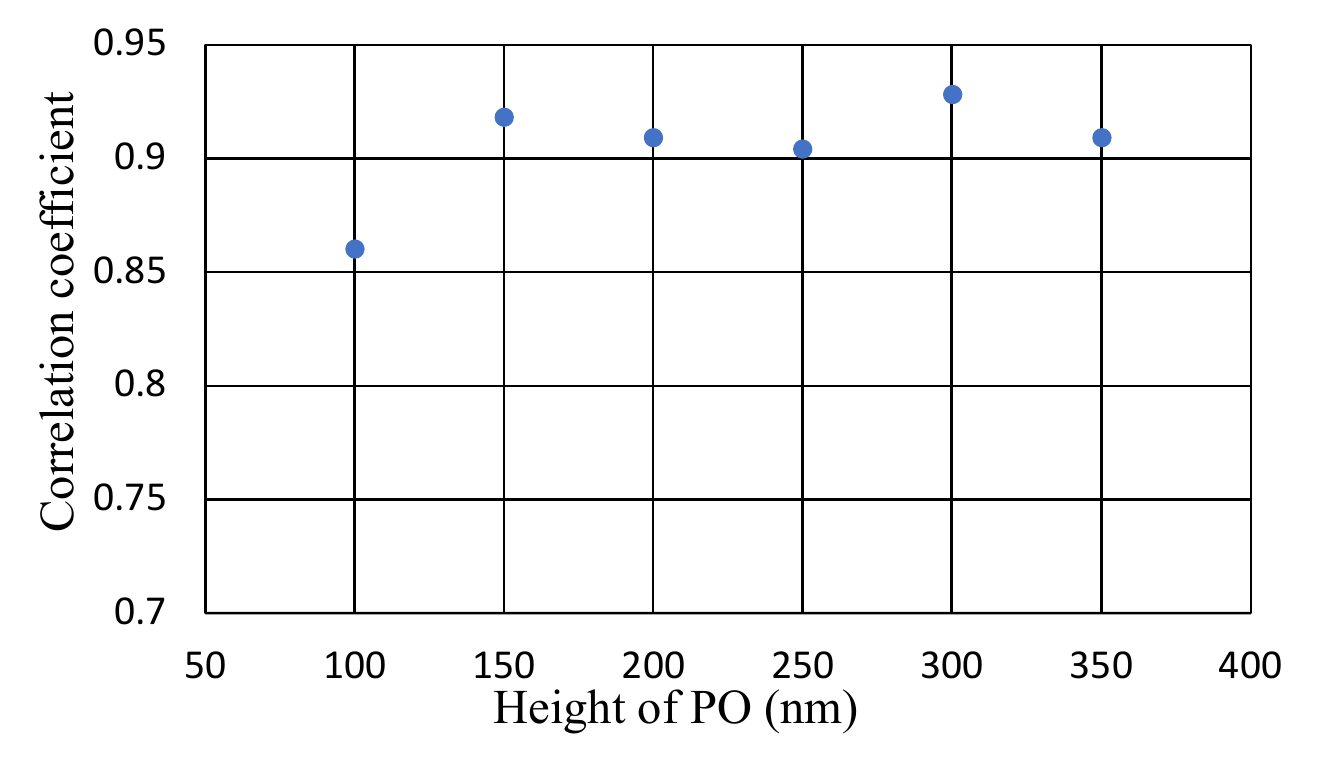}
    \caption{Correlation between ideal edge enhancement and edge enhancement for PO for different heights.}
    \label{4exp}
\end{figure}
In an effort to quantify the edge enhancement, we used the same ideal edge from Fig. \ref{1exp} (c) to compare with the edge enhancement of PO 350-nm, because for the same function and dimension of AO and PO, they produce the same edge enhancement. Figs. \ref{2exp} (c) and (d) are extracted from the wrapped square portion of Figs. \ref{2exp} (a) and (b), respectively. The horizontal cross-sectional intensity distribution in Fig. \ref{2exp} (e) is plotted along the white dashed line of Fig. \ref{2exp} (c) (green curve), Fig. \ref{2exp} (d) (blue curve), and ideal edge of Fig. \ref{1exp} (c) (red curve). Figure \ref{2exp} (e) shows the edge of the 350-nm-height PO without q-plate (green curve) is not enhanced clearly compared to blue curve which used a q-plate.  

The Fig. \ref{4exp} shows the correlation coefficient between the ideal edge Fig. \ref{1exp} (c) and PO for different heights. As shown in Fig. \ref{4exp}, the correlation coefficient increases as the height increases from 100 to 350 nm, implying that the edge enhancement is better at 350 nm (correlation coefficient: 0.909) than that at 100 nm (correlation coefficient: 0.860). This is because the phase difference at a height of 100 nm, which is 0.56 rad, is lower than the phase difference at 350 nm, which is 1.99 rad. The lower the phase difference, flatter is the slope/gradient of the object, and hence, the $4f$ system does not enhance the edge clearly. 

\subsection{Observation of an onion cell}
\begin{figure}[ht]
    \centering
    \includegraphics[width=0.80\linewidth]{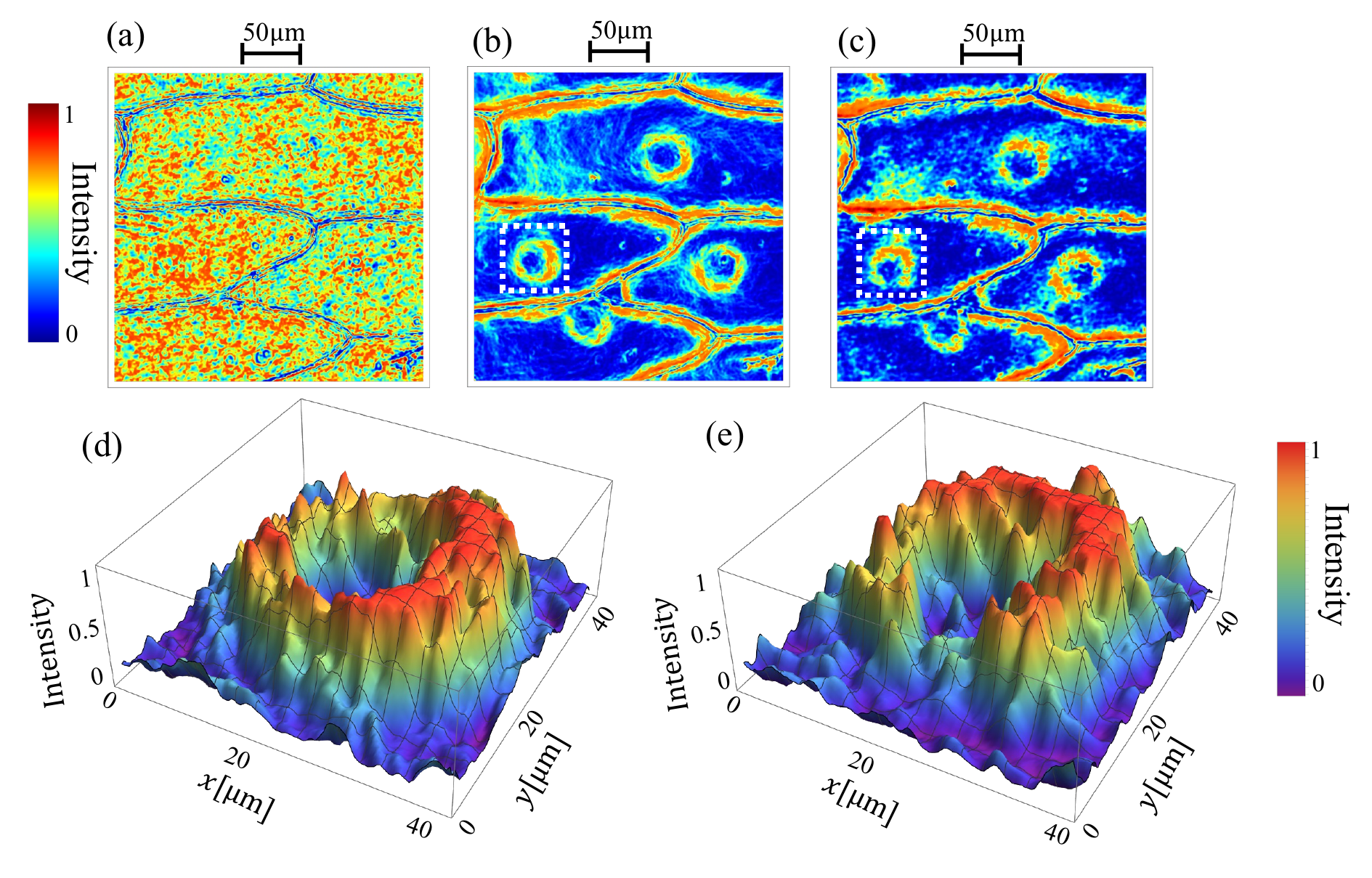}
    \caption{Experimental results of onion cell, (a) without q-plate, (b) VVF image, (c) SVF image, and (d) and (e) surface plot of the phase edge of wrapped portion in Figs. \ref{fig6} (b) and (c).} 
    \label{fig6}
\end{figure}
 SVF is capable of enhancing the edges of AOs and POs. However, as per our calculation, Eq. (\ref{Eq6}), if a sample is a complex PAO, it enhances or reduces the edge intensity due to third term on the right hand side of Eq. (\ref{Eq6}), which is absent in the case of VVF. Now, in order to verify our theoretical calculation, we changed the sample to an onion cell, which contains a complex PAO. The transparent cell nucleus and an oblique cell wall in onion cell are the PO and AO, respectively.  Fig. \ref{fig6} (a) shows the observation results of onion cells without a q-plate. Figs. \ref{fig6} (b) and (c) show the edge enhancement using VVF and SVF, respectively. The cell wall, which is the AO, can be seen without the q-plate in Fig. \ref{fig6} (a), but the edge of the cell nucleus, which is the PO, are difficult to see. However, in Figs. \ref{fig6} (b) and (c), both the edges of the cell wall and cell nucleus can be seen after inserting the q-plate. Due to the absolute square value of the partial derivative on the cell wall, we can see two lines on the cell wall as the edge in Figs. \ref{fig6} (b) and (c). A surface plot in Figs. \ref{fig6} (d) and (e) shows the phase edge profile from the portion wrapped with white dashed square in Figs. \ref{fig6} (b) and (c). The surface plot indicates that the intensity of the phase edge is enhanced by a $4f $system with a q-plate, because the intensity is higher than that of the surrounding area. If we compare two surface plots,  we can see the highest intensity is almost continuous in Fig. \ref{fig6} (d) compared to that is Fig. \ref{fig6} (e), in which the highest intensity is broken at the bottom side. The discontinuity of the edges in SVF image is because of the contribution of the positive and negative gradient from the third term on the right hand side of Eq. \ref{Eq6}. In order to support our theory and experimental results, we used the numerical simulation described in Section \ref{D1section}.

\section{Discussion}
In the first part, we will discuss the simulation result obtained to support our theoretical and experimental results on the onion cell. Then, in the second part, we will demonstrate the separation of the PO edge from the AO edge through theoretical calculation and numerical simulation.

\subsection{Evidence to support our theoretical calculation and experimental results on the onion cell}
\label{D1section}
 We prepared a sample similar to an onion cell for simulation. In order to set up simulation conditions,  we selected the portion, wrapped in white dashed line of Fig. \ref{fig6} (b) and collected the information. The radius of the cell nucleus, thickness of the cell wall, and distance from the centre of cell nucleus to the centre of the cell wall were 14, 5, and 27 \textmu m , respectively. With this information, we prepared a PO with radius 14 \textmu m using a super Gaussian function of order 5. Similarly, an AO with thickness 5 \textmu m  using a one dimensional super Gaussian function of order 5 was prepared. Then, the PO was placed 27 \textmu m  away from the AO to make the PAO.
 
\begin{figure}[ht]
    \centering
    \includegraphics[width=0.60\linewidth]{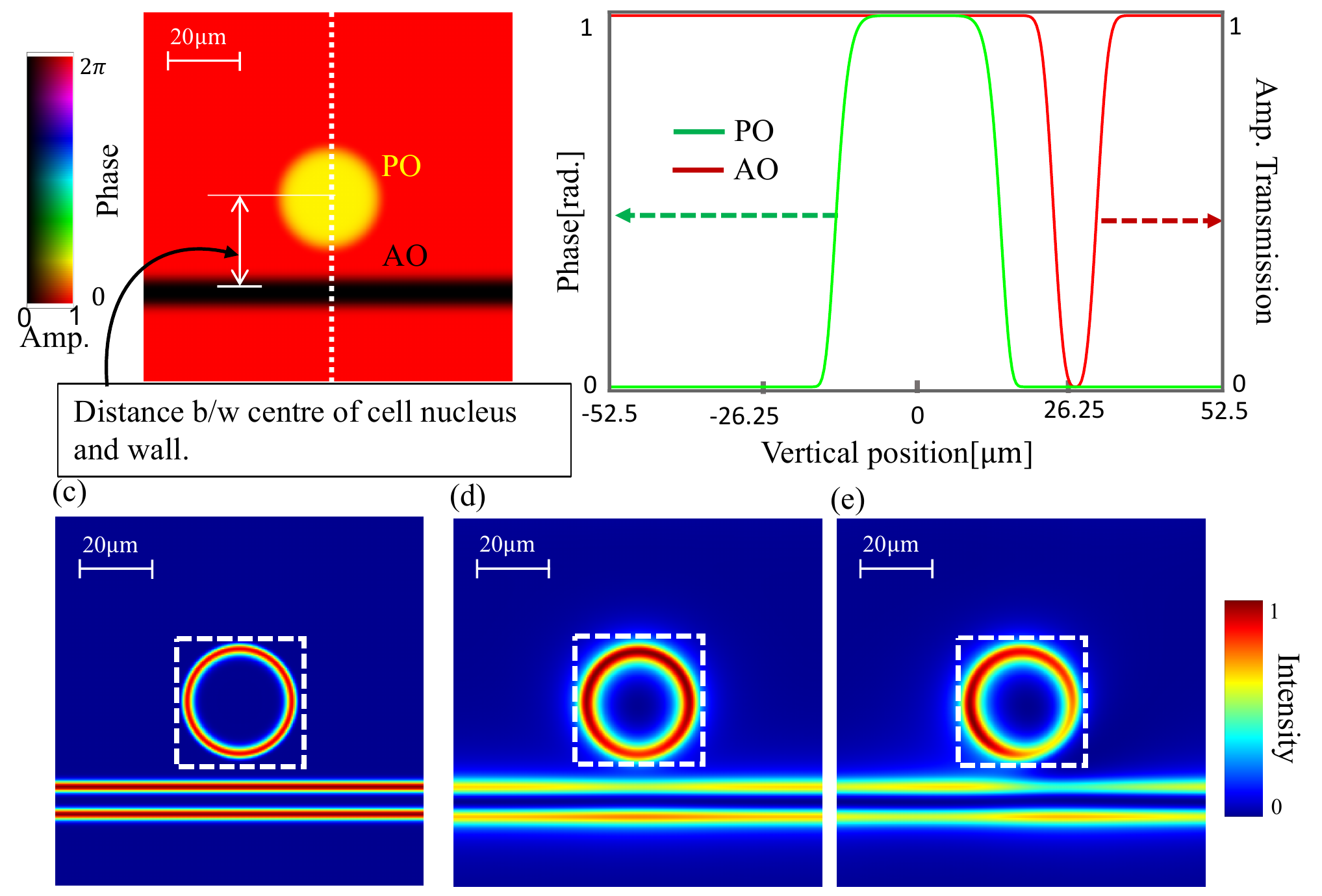}
    \caption{(a) Distribution of PAO, (b) vertical cross-sectional distributions of the PAO, (c) ideal edge enhancement of PAO, (d) edge enhancement of PAO with VVF, and (e) edge enhancement of PAO with SVF.}
    \label{D1}
\end{figure}

 Figs. \ref{D1} (a) and (b) shows the PAO and its vertical cross-sectional distributions, respectively. The ideal edge enhancement of PAO is shown in Fig. \ref{D1} (c). Figs. \ref{D1} (d) and (e) represent edge enhancement of the PAO using VVF and SVF, respectively. 

\begin{figure}[ht]
    \centering
    \includegraphics[width=0.55\linewidth]{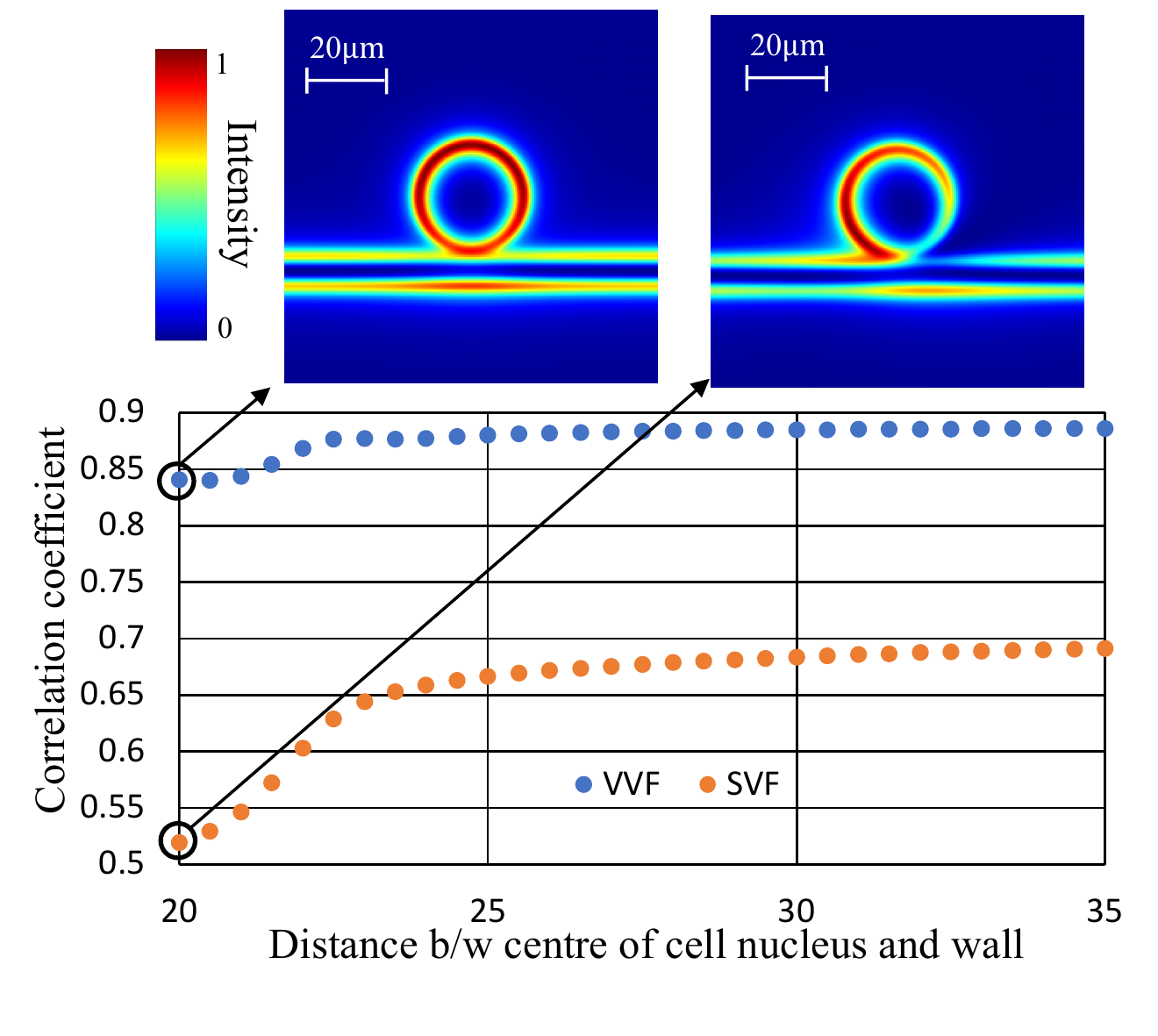}
    \caption{Correlation coefficient: Blue scatter shows correlation between ideal PAO edge and PAO edges obtained by VVF, and orange scatter shows correlation between ideal PAO edge and PAO edges obtained by SVF.}
    \label{D2}
\end{figure}

 It is obvious from Fig. \ref{D1} (d) that the edge of the PO is continuously highlighted compared to Fig. \ref{D1} (e), where the intensity decreases at the bottom-right hand side compared to that on the left hand side. This is because as per Eq. (\ref{Eq6}), the extra third term from the negative gradient, $\left( \partial_x B(r)\partial_y A(r)-\partial_x A(r)\partial_y B(r) \right)<0$, reduces the intensity from the first and second terms. 

In order to determine the correlation coefficient, only the cell nucleus were selected as shown in Fig. \ref{D1} (wrapped with white dotted square for ideal, VVF, and SVF images). The blue scatter in Fig. \ref{D2} represent the correlation coefficient between the ideal PAO edge and PAO edges obtained by VVF for varying distance between the cell nucleus and cell wall. Similarly, the orange scatter shows the correlation between the ideal PAO edge and PAO edges obtained by SVF for varying distance between the cell nucleus and cell wall. At 20 \textmu m, i.e., the distance between the center of the cell nucleus and wall, the correlation coefficient is minimum for both VVF and SVF. This is because at the minimum distance, the edges of the cell nucleus and wall overlap, as can be seen in Fig. \ref{D2}. However, the correlation coefficient remains approximately the same for VVF (slope:0.01), whereas the correlation coefficient increases as the distance increases for SVF (slope:0.04). This indicates that as the cell nucleus (PO) moves away from cell wall (AO), the interference from AO to PO is minimized.

From the experimental results of the onion cell and simulation result of VVF images, the edges of the PO and AO present in the PAO are enhanced equally, making it difficult to distinguish the two. Therefore, we present a numerical calculation and simulation to eliminate the AO edge from that of the PAO in the next section.

\subsection{Isolation of PO from AO}
\label{Isolation of Phase Object from Amplitude-phase object} 
\begin{figure}[htbp]
    \centering
    \includegraphics[width=0.60\linewidth]{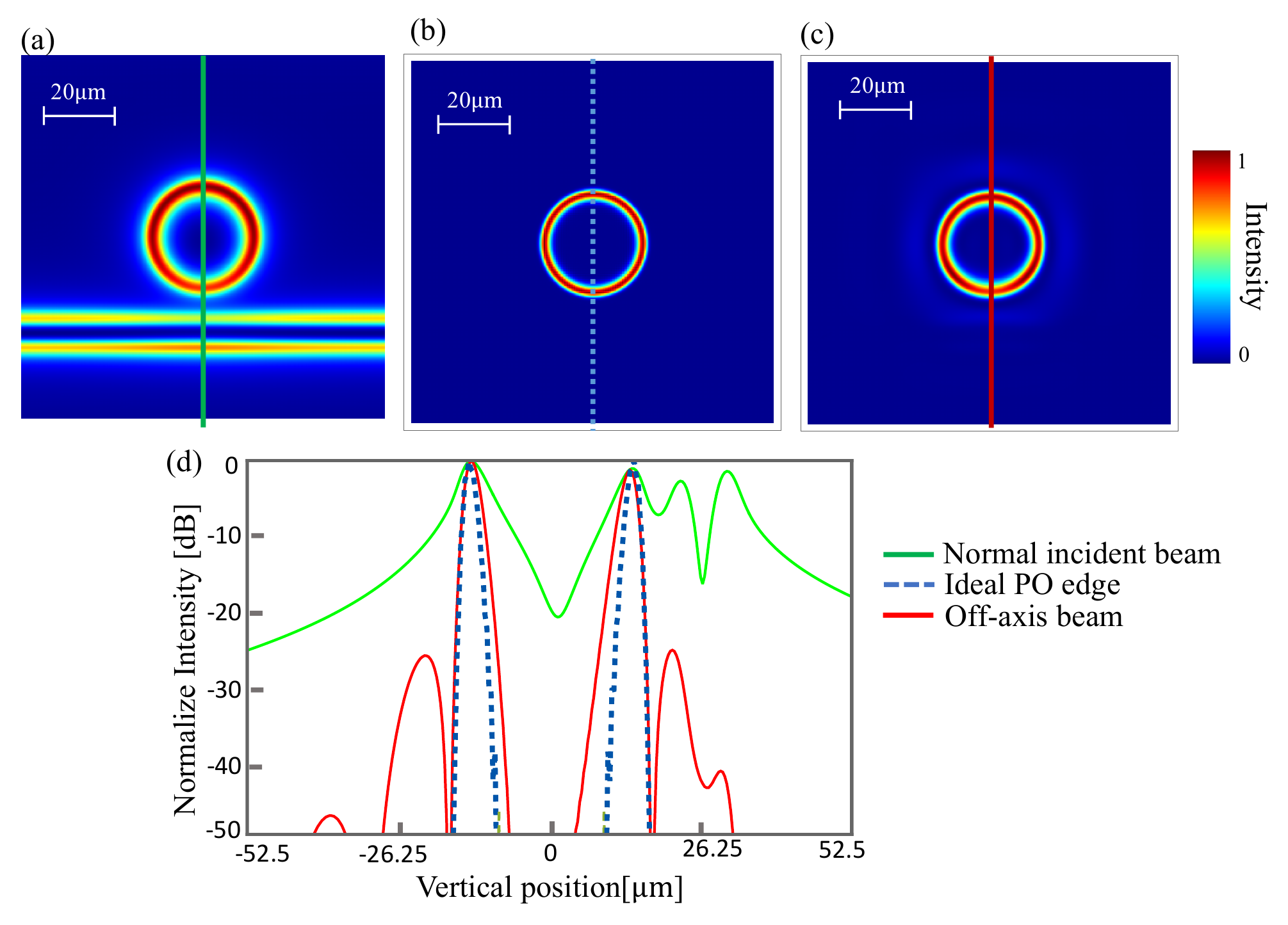}
    \caption{(a) Edge enhancement of PAO with normal incident beam, (b) ideal edge enhancement of the PO, (c) isolated edge enhancement of the PO with off-axis beam, and (d) vertical cross-sectional logarithmic intensity distributions for ideal PO edge, normal incident beam, and off-axis beam.}
    \label{ISO1}
\end{figure}
In order to eliminate the AO from PAO, we use an off-axis beam, which deviates from the propagation direction $z$. The deviated beam on the $xz$ and $yz$ plane with a small angle, $\beta \ll 1$, gives extra phase on the input plane wave as $\exp\left(\ii \beta\bm{k}\cdot\bm{r}\right)$, where $\bm{k}=(k_x,k_y)=\beta k\bm{e}$ is the two-dimensional transverse wavenumber vector with $\bm{e}$ being a two-dimensional unit vector describing the transverse component of propagation direction. The extra phase $\exp\left(\ii\beta\bm{k}\cdot\bm{r}\right)$ will contributes to PO edge at the output intensity, $I$, as shown below.
\begin{equation}
    I(\bm{r},\bm{e},\beta)\approx\left|\bm{\nabla}\left[f_{\text{in}} (\bm{r})\ee^{\ii\beta k\bm{e}\cdot \bm{r}}\right]\right|^2
    =\left|\bm{\nabla}A(\bm{r})\right|^2
    +A(\bm{r})^2\left|\bm{\nabla}B(\bm{r})+\beta k\bm{e}\right|^2.
    \label{Eq9}
\end{equation}
The first and second terms on the right hand side of Eq. (\ref{Eq9}) correspond to the AO edge, and the PO edge with the contribution of the off-axis beam, respectively. 
Since only the second term contains the contribution of the off-axis beam in Eq. (\ref{Eq9}), the first term (AO edge) can be eliminated by the following calculation for four light waves with different propagational directions:
\begin{equation}
\begin{split}
    J(\bm{r},\beta)&\equiv\left\{
        I(\bm{r},\bm{e}_x,\beta)-I(\bm{r},\bm{e}_x,-\beta)
        \right\}^2
        +\left\{
        I(\bm{r},\bm{e}_y,\beta)-I(\bm{r},\bm{e}_y,-\beta)
        \right\}^2\\
    &\approx \left|4\beta k A(\bm{r})^2\bm{\nabla}B(\bm{r})\right|^2,
    \end{split}
    \label{Eq10}
\end{equation}
where $\bm{e}_x$ and $\bm{e}_y$ are unit vectors along the $x$ and $y$ axes, respectively.
Two pairs of off-axis beams with positive and negative angles along $\bm{e}_x$ and $\bm{e}_y$ can generate $x$- and $y$-directional edge detection of the PO, and they are combined to produce two-dimensional edge enhancement, resulting in isolation of the PO edge from the PAO.

As a numerical demonstration, we considered the same sample described in Section \ref{D1section} and Fig. \ref{D1} (a), except the distance between cell nucleus and cell wall was reduced from 27 \textmu m to 25 \textmu m. This is because we want the cell nucleus and cell wall to be sufficiently close.  
The edge enhancement with normal incident beam, $\beta=0\degree$, and ideally isolated edge-enhanced image of the PO is shown in Figs. \ref{ISO1} (a) and (b), respectively. The AO edge observed at the normal incident light as shown in Fig. \ref{ISO1} (a) could be eliminated according to Eq. \ref{Eq10}. The isolated edge enhancement of the PO from AO using an off-axis beam is shown in Fig. \ref{ISO1} (c). Here, we assumed the off-axis beam $\beta=0.1\degree$ and wavelength $\lambda=2\pi/k=635$ nm.
As can be seen from the logarithmic intensity of the horizontal cross-sectional distribution shown in Fig. \ref{ISO1} (c), the AO edge was eliminated using the off-axis beam case (red curve) compared to the normal incident (green curve). The minimum intensity ratio between the AO edge that uses the normal incident beam (green curve) and off-axis beam (red curve) is -23dB, which is suitable for edge reduction. 

\section{Conclusion}
In this paper, edge enhancement of a PAO using a $4f$ system with a q-plate was calculated theoretically and verified experimentally. The q-plate filter generates SVF or VVF by using different polarization states of the illumination beam. The experimental results of an edge enhanced onion cell using SVF and VVF indicate the validity of the theory and method. We quantitatively compare between the cell nucleus edge between SVF and VVF images. Furthermore, our numerical simulation results obtained at similar conditions as for the experimental setup agree well with the experimental results. We also proposed a new method to isolate the PO edge from the AO edge using off-axis beam illumination theoretically, and verified it via numerical simulation. The proposed isolation of PO edge could be used in microscopy research or biological edge detection, and we will conduct a proof-of-principle experiment in a future study.

\section*{Funding}
This work has been partially supported by the Research Foundation for Opt-Science and Technology and partially by JSPS KAKENHI (grant no. 20K05364 and 18KK0079).

\section*{Disclosures}
The authors declare no conflicts of interest.

\bibliography{references} 

\begin{thebibliography}{10}
\newcommand{\enquote}[1]{``#1''}

\bibitem{wang2012detecting}
G.~Wang and N.~Fang, \enquote{Detecting and tracking nonfluorescent
  nanoparticle probes in live cells,} {\protect\JournalTitle{Methods in
  Enzymology}} \textbf{504}, 83--108 (2012).

\bibitem{khandpur2020compendium}
R.~S. Khandpur, \emph{Compendium of Biomedical Instrumentation, 3 Volume Set}
  (John Wiley \& Sons, 2020).

\bibitem{burch1942phase}
C.~Burch and J.~Stock, \enquote{Phase-contrast microscopy,}
  {\protect\JournalTitle{Journal of Scientific Instruments}} \textbf{19}, 71
  (1942).

\bibitem{frohlich2008phase}
V.~C. Frohlich, \enquote{Phase contrast and differential interference contrast
  ({DIC}) microscopy,} {\protect\JournalTitle{JoVE (Journal of Visualized
  Experiments)}} p. e844 (2008).

\bibitem{lang1982nomarski}
W.~Lang, \emph{Nomarski differential interference-contrast microscopy} (Carl
  Zeiss Oberkochen, 1982).

\bibitem{5}
Y.~Zhou, S.~Feng, S.~Nie, J.~Ma, and C.~Yuan, \enquote{Image edge enhancement
  using airy spiral phase filter,} {\protect\JournalTitle{Optics Express}}
  \textbf{24}, 25258--25268 (2016).

\bibitem{9}
M.~K. Sharma, J.~Joseph, and P.~Senthilkumaran, \enquote{Selective edge
  enhancement using shifted anisotropic vortex filter,}
  {\protect\JournalTitle{Journal of Optics}} \textbf{42}, 1--7 (2013).

\bibitem{11}
J.~Wang, W.~Zhang, Q.~Qi, S.~Zheng, and L.~Chen, \enquote{Gradual edge
  enhancement in spiral phase contrast imaging with fractional vortex filters,}
  {\protect\JournalTitle{Scientific Reports}} \textbf{5}, 1--6 (2015).

\bibitem{imageprocessing1}
Z.~Gu, D.~Yin, S.~Nie, S.~Feng, F.~Xing, J.~Ma, and C.~Yuan,
  \enquote{High-contrast anisotropic edge enhancement free of shadow effect,}
  {\protect\JournalTitle{Applied Optics}} \textbf{58}, G351--G357 (2019).

\bibitem{imageprocessing2}
C.-S. Guo, Y.-J. Han, J.-B. Xu, and J.~Ding, \enquote{Radial {H}ilbert
  transform with {L}aguerre-{G}aussian spatial filters,}
  {\protect\JournalTitle{Optics Letters}} \textbf{31}, 1394--1396 (2006).

\bibitem{biological1}
S.~Wei, S.~Zhu, and X.~Yuan, \enquote{Image edge enhancement in optical
  microscopy with a bessel-like amplitude modulated spiral phase filter,}
  {\protect\JournalTitle{Journal of Optics}} \textbf{13}, 105704 (2011).

\bibitem{biological2}
Y.~Zhou, S.~Feng, S.~Nie, J.~Ma, and C.~Yuan, \enquote{Anisotropic edge
  enhancement with spiral zone plate under femtosecond laser illumination,}
  {\protect\JournalTitle{Applied Optics}} \textbf{56}, 2641--2648 (2017).

\bibitem{biological3}
G.~Situ, M.~Warber, G.~Pedrini, and W.~Osten, \enquote{Phase contrast
  enhancement in microscopy using spiral phase filtering,}
  {\protect\JournalTitle{Optics Communications}} \textbf{283}, 1273--1277
  (2010).

\bibitem{medical}
P.~Shankar, \enquote{Contrast enhancement and phase-sensitive boundary
  detection in ultrasonic speckle using bessel spatial filters,}
  {\protect\JournalTitle{IET Image Processing}} \textbf{3}, 41--51 (2009).

\bibitem{fingerprint1}
M.~K. Sharma, J.~Joseph, and P.~Senthilkumaran, \enquote{Directional edge
  enhancement using superposed vortex filter,} {\protect\JournalTitle{Optics \&
  Laser Technology}} \textbf{57}, 230--235 (2014).

\bibitem{28}
K.~Kohlmann, \enquote{Corner detection in natural images based on the 2-{D}
  {H}ilbert transform,} {\protect\JournalTitle{Signal Processing}} \textbf{48},
  225--234 (1996).

\bibitem{29}
S.~Khonina, V.~Kotlyar, M.~Shinkaryev, V.~Soifer, and G.~Uspleniev,
  \enquote{The phase rotor filter,} {\protect\JournalTitle{Journal of Modern
  Optics}} \textbf{39}, 1147--1154 (1992).

\bibitem{30}
A.~Papoulis, \enquote{Optical systems, singularity functions, complex hankel
  transforms,} {\protect\JournalTitle{JOSA}} \textbf{57}, 207--213 (1967).

\bibitem{31}
Y.-J. Han, C.-S. Guo, Z.-Y. Rong, and L.-M. Chen, \enquote{Radial {H}ilbert
  transform with the spatially variable half-wave plate,}
  {\protect\JournalTitle{Optics Letters}} \textbf{38}, 5169--5171 (2013).

\bibitem{32}
J.~A. Davis, D.~E. McNamara, D.~M. Cottrell, and J.~Campos, \enquote{Image
  processing with the radial {H}ilbert transform: theory and experiments,}
  {\protect\JournalTitle{Optics Letters}} \textbf{25}, 99--101 (2000).

\bibitem{spiralPF1}
Z.~Li, S.~Zhao, and L.~Wang, \enquote{Isotropic and anisotropic edge
  enhancement with a superposed-spiral phase filter,}
  {\protect\JournalTitle{Optics Express}} \textbf{29}, 32591--32602 (2021).

\bibitem{spiralPF2}
S.~F{\"u}rhapter, A.~Jesacher, S.~Bernet, and M.~Ritsch-Marte, \enquote{Spiral
  phase contrast imaging in microscopy,} {\protect\JournalTitle{Optics
  Express}} \textbf{13}, 689--694 (2005).

\bibitem{spiralPF3}
R.~Juchtmans, L.~Clark, A.~Lubk, and J.~Verbeeck, \enquote{Spiral phase plate
  contrast in optical and electron microscopy,} {\protect\JournalTitle{Physical
  Review A}} \textbf{94}, 023838 (2016).

\bibitem{spiralPF4}
A.~Jesacher, S.~F{\"u}rhapter, S.~Bernet, and M.~Ritsch-Marte, \enquote{Shadow
  effects in spiral phase contrast microscopy,} {\protect\JournalTitle{Physical
  Review Letters}} \textbf{94}, 233902 (2005).

\bibitem{spiralPF5}
X.~Qiu, F.~Li, W.~Zhang, Z.~Zhu, and L.~Chen, \enquote{Spiral phase contrast
  imaging in nonlinear optics: seeing phase objects using invisible
  illumination,} {\protect\JournalTitle{Optica}} \textbf{5}, 208--212 (2018).

\bibitem{spiralPF6}
G.~Situ, G.~Pedrini, and W.~Osten, \enquote{Spiral phase filtering and
  orientation-selective edge detection/enhancement,}
  {\protect\JournalTitle{JOSA A}} \textbf{26}, 1788--1797 (2009).

\bibitem{spiralPF7}
S.~Bernet, A.~Jesacher, S.~F{\"u}rhapter, C.~Maurer, and M.~Ritsch-Marte,
  \enquote{Quantitative imaging of complex samples by spiral phase contrast
  microscopy,} {\protect\JournalTitle{Optics Express}} \textbf{14}, 3792--3805
  (2006).

\bibitem{spiralPF8}
S.~F{\"u}rhapter, A.~Jesacher, C.~Maurer, S.~Bernet, and M.~Ritsch-Marte,
  \enquote{Spiral phase microscopy,} {\protect\JournalTitle{Advances in Imaging
  and Electron Physics}} \textbf{146}, 1--59e (2007).

\bibitem{33}
Y.~Kim, G.-Y. Lee, J.~Sung, J.~Jang, and B.~Lee, \enquote{Spiral metalens for
  phase contrast imaging,} {\protect\JournalTitle{Advanced Functional
  Materials}} \textbf{32}, 2106050 (2022).

\bibitem{RPC1}
X.~Qiu, F.~Li, W.~Zhang, Z.~Zhu, and L.~Chen, \enquote{Spiral phase contrast
  imaging in nonlinear optics: seeing phase objects using invisible
  illumination,} {\protect\JournalTitle{Optica}} \textbf{5}, 208--212 (2018).

\bibitem{RPC2}
L.~Liu, H.~Wang, Y.~Ning, C.~Guo, and G.~Ren, \enquote{Infrared upconverted
  image edge enhancement using spiral phase filter,}
  {\protect\JournalTitle{Laser Physics}} \textbf{29}, 015401 (2018).

\bibitem{qplate3}
R.~Gozali, T.-A. Nguyen, E.~Bendau, and R.~R. Alfano, \enquote{Compact {OAM}
  microscope for edge enhancement of biomedical and object samples,}
  {\protect\JournalTitle{Review of Scientific Instrum.}} \textbf{88}, 093701
  (2017).

\bibitem{qplate1}
D.~Xu, T.~Ma, X.~Qiu, W.~Zhang, and L.~Chen, \enquote{Implementing selective
  edge enhancement in nonlinear optics,} {\protect\JournalTitle{Optics
  Express}} \textbf{28}, 32377--32385 (2020).

\bibitem{qplate2}
D.~Li, S.~Feng, S.~Nie, J.~Ma, and C.~Yuan, \enquote{Scalar and vectorial
  vortex filtering based on geometric phase modulation with a q-plate,}
  {\protect\JournalTitle{Journal of Optics}} \textbf{21}, 065702 (2019).

\bibitem{swaveplate1}
B.~Zhang, Z.~Chen, H.~Sun, J.~Xia, and J.~Ding, \enquote{Vectorial optical
  vortex filtering for edge enhancement,} {\protect\JournalTitle{Journal of
  Optics}} \textbf{18}, 035703 (2016).

\bibitem{swaveplate2}
B.~B. Ram, P.~Senthilkumaran, and A.~Sharma, \enquote{Polarization-based
  spatial filtering for directional and nondirectional edge enhancement using
  an s-waveplate,} {\protect\JournalTitle{Applied Optics}} \textbf{56},
  3171--3178 (2017).

\bibitem{34}
L.~Marrucci, \enquote{The q-plate and its future,}
  {\protect\JournalTitle{Journal of Nanophotonics}} \textbf{7}, 078598 (2013).

\bibitem{35}
A.~Rubano, F.~Cardano, B.~Piccirillo, and L.~Marrucci, \enquote{Q-plate
  technology: a progress review,} {\protect\JournalTitle{JOSA B}} \textbf{36},
  D70--D87 (2019).

\bibitem{36}
S.~Franke-Arnold and N.~Radwell, \enquote{Light served with a twist,}
  {\protect\JournalTitle{Optics and Photonics News}} \textbf{28}, 28--35
  (2017).

\bibitem{37}
J.~A. Davis, N.~Hashimoto, M.~Kurihara, E.~Hurtado, M.~Pierce, M.~M.
  S{\'a}nchez-L{\'o}pez, K.~Badham, and I.~Moreno, \enquote{Analysis of a
  segmented q-plate tunable retarder for the generation of first-order vector
  beams,} {\protect\JournalTitle{Applied Optics}} \textbf{54}, 9583--9590
  (2015).

\bibitem{38}
S.~Delaney, M.~M. S{\'a}nchez-L{\'o}pez, I.~Moreno, and J.~A. Davis,
  \enquote{Arithmetic with q-plates,} {\protect\JournalTitle{Applied Optics}}
  \textbf{56}, 596--600 (2017).

\end{thebibliography}
\end{document}